\documentstyle[11pt,moriond,epsfig]{article}

\def\be{\begin{equation}}
\def\ee{\end{equation}}
\def\bea{\begin{eqnarray}}
\def\eea{\end{eqnarray}}

\begin{document}
\vspace*{4cm}
\title{A (RE)INTERPRETATION OF THE QCD PHASE TRANSITION AND OF 
STRANGENESS AS QGP SIGNATURE}

\author{SONJA KABANA}

\address{Laboratory for High Energy Physics, University of Bern,
Sidlerstrasse 5, 3012 Bern, Switzerland}

\maketitle\abstracts{
The temperature
at the chemical freeze-out and at zero baryochemical potential
has been extracted 
in a global analysis of $e^+e^-$, $p+p$, $p+ \overline{p}$ and
$A+A$ collisions at $\sqrt{s}$=2-1800 GeV per N+N pair.
We demonstrate that 
the temperature at $\mu_B$=0, rises with the initial energy density 
$\epsilon_i$,
and saturates above $\epsilon_i$ $\sim$ 1 GeV/fm$^3$.
This behaviour is interpreted as mapping out the QCD phase transition
universally in particle and nuclear collisions.
The critical energy density is therefore identified to be
 $\epsilon_{crit}$ $\sim$ 1 $\pm$ 0.3 GeV/fm$^3$.
We  show that strange particles at $\mu_B$=0, are not
significantly enhanced in A+A collisions as compared to $p+ \overline{p}$.
The so called 
'strangeness suppression factor' ($\lambda_s = \frac{ (2 \overline{s}) }
{ (\overline{u} + \overline{d}) }$) as a function of $\epsilon_i$
is following the temperature,
rising and saturating universally above $\epsilon_{crit}$.
This leads to a  reinterpretation of strangeness enhancement as
QGP signature.
Within this interpretation the experimental puzzles with respect to
strangeness production can be naturally explained:
e.g.  the recent measured maximum of $K^+/\pi^+$ in
Pb+Pb collisions at 40 A GeV, is explained as due to $\mu_B$. 
We discuss under which conditions  'strangeness enhancement'
and  '$J/\Psi$ suppresion' 
 both set in at $\epsilon_{crit}$ $\sim$ 1 GeV/fm$^3$.
}

\section{Introduction}\label{subsec:intro}

\noindent
One outstanding prediction of the theory of strong interaction
(e.g. \cite{lattice,mapping}) is the phase transition from
confined hadrons to a deconfined phase of their constituents, the
quarks and gluons (the so called quark gluon plasma =QGP).
An experimental program which started approximately in the
eighties and continues
in many accelerators as: CERN SPS, BNL  RHIC, CERN LHC and GSI SIS,
has been dedicated to the experimental verification of this
transition.
\\
\noindent
There is at present evidence that the QCD phase transition
occurs in central Pb+Pb reactions at 158 A GeV \cite{heinzjakob}.
The major evidence \cite{heinzjakob} 
is 1) the suppression of the $J/\Psi/DY$ ratio
 \cite{na50}, 2) the enhancement of
strange particles and in particular of strange antibaryons
 \cite{strangeness}
and 3) enhancement of $\gamma$, $\mu^+ \mu^-$ and $e^+ e^-$ 
production \cite{wa98,na50_mm,na45,na45_qm2001}.
New results from RHIC added a fourth item, namely indications of
jet quenching \cite{bjacak_this_confernece}.
However, many aspects of the experimental data
remain theoretically unclear and are even selfcontradicting:

\noindent
(1) An example of unclear interpretation concerns
 the pattern of $J/\Psi/DY$ suppression. For example it can be described
by one \cite{olli} or two suppression steps \cite{satz},
or  within approaches without QGP formation \cite{capella}.
\\
(2) An example of selfcontradiction is strange particle production.
Strange particle enhancement 
is considered by several authors as QGP signature, at and above
 SPS energy \cite{rafelski}.
However this enhancement (when defined as double ratio of: $K/\pi$(A+A/p+p)),
 increases with decreasing $\sqrt{s}$ \cite{ogilvie}.
This behaviour is the opposite as expected, as the
transition should occur in the highest $\epsilon_i$ and not in the lowest.
Other
 authors \cite{marek} conclude from the latter observation, 
that it seems --phenomenologically-- better to compare the
$\sqrt{s}$ dependence of $K/\pi$ in A+A reactions only, without
comparing to p+p. 
They also propose that the $K/\pi$ ratio would signal the transition,
through a sudden drop towards the higher $\sqrt{s}$,  at the critical
 $\sqrt{s}$. 
Therefore the  QGP signature would be
  'strangeness suppression' --towards the higher $\sqrt{s}$ -- 
and not 'enhancement'.
Such a drop has indeed been observed in
Pb+Pb collisions between 40 and 158 A GeV \cite{na49}.
Some
 authors (e.g. \cite{redlich_qm2001}) conclude that the data on 
strange particle production agree but do not prove QGP formation.
 Another
 puzzling observation is that strange particles are
 enhanced also in p+A collisions as compared to p+p \cite{phi}\footnote{
However
the enhancement seen in the $\phi/\pi$ ratio in p+A over p+p collisions
at 158 GeV per nucleon 
by NA49 \cite{phi} is restricted in the forward rapidity region,
while this ratio has not yet been compared at midrapidity and at full
acceptance.}
 --puzzling because
of the implied assumption that the QCD phase transition does not occur
in p+p, p+A and peripheral A+A reactions.
\\
Indeed, mostly, all
 signatures of the QGP are extracted comparing central A+A reactions
mainly to 
  p+p, p+A and peripheral A+A collisions and to models
which have been shown to reproduce the latter reactions without
invoking QGP formation.
This comparison implies the assumption that no QGP can be formed
in the latter reactions. This assumption is however a priori unproven.

\vspace{0.3cm}

\noindent
In this paper we address the following questions:
\\
a) Is there a simultaneous appearance of all QGP signatures ?
We will mainly address the two QGP signatures  of
'$c \overline{c}$ suppression' and '$s$ and $ \overline{s}$ enhancement' 
\cite{0004138}, because of lack of measurements of the other
signatures at low $\epsilon_i$.
\\
b) Is there evidence for the QCD phase transition
in global observables showing a discontinuity,
beyond rare probes as the $J/\Psi$ or the $\Omega$ data ?
\cite{0010228,mapping}.
\\
c)
Can we extract the critical energy density
($\epsilon_{crit}$) above which the
QCD phase transition occurs?
\cite{mapping}.
\\
d) Is the initial energy density ($\epsilon_{i}$) the only discriminating
parameter for this transition, or
 is there in addition a 'critical' volume
which must be reached --to achieve equilibrium--  ?
In other words, can the QGP phase transition occur in 
reactions of leptons and/or hadrons or only in nuclear reactions ? 
\cite{0010228,mapping}.
\\
e) Can we explain the above mentioned puzzles (2) which concern
 the production of strange particles ?
Is strangeness a QGP signature ?
If so, is strangeness enhancement or suppression a QGP signature ?
\cite{border}.
\\
We follow these questions one by one in the subsequent sections.

\begin{figure}
\begin{center}
\psfig{figure=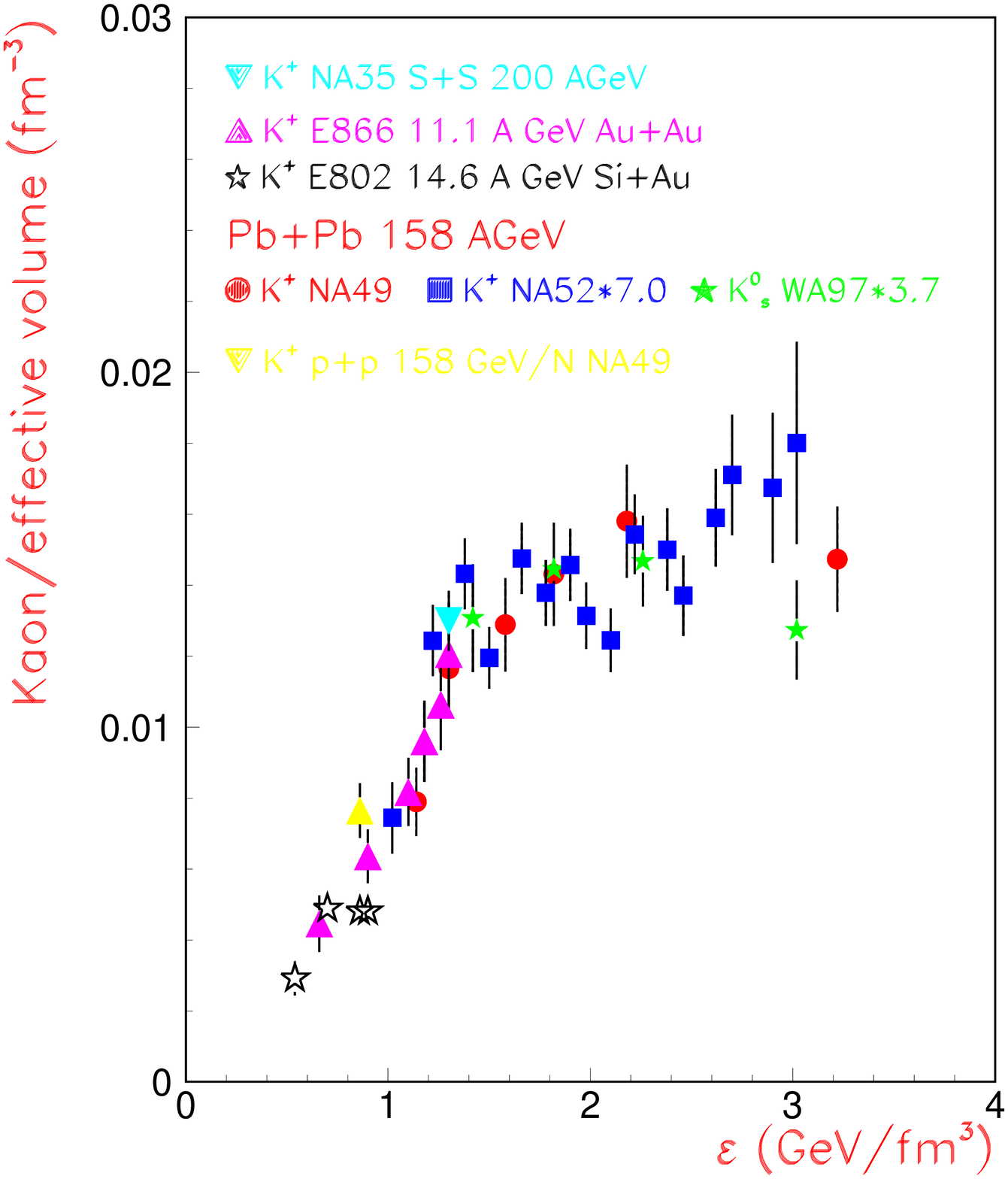,height=3.2in}
\hspace*{0.3cm}
\psfig{figure=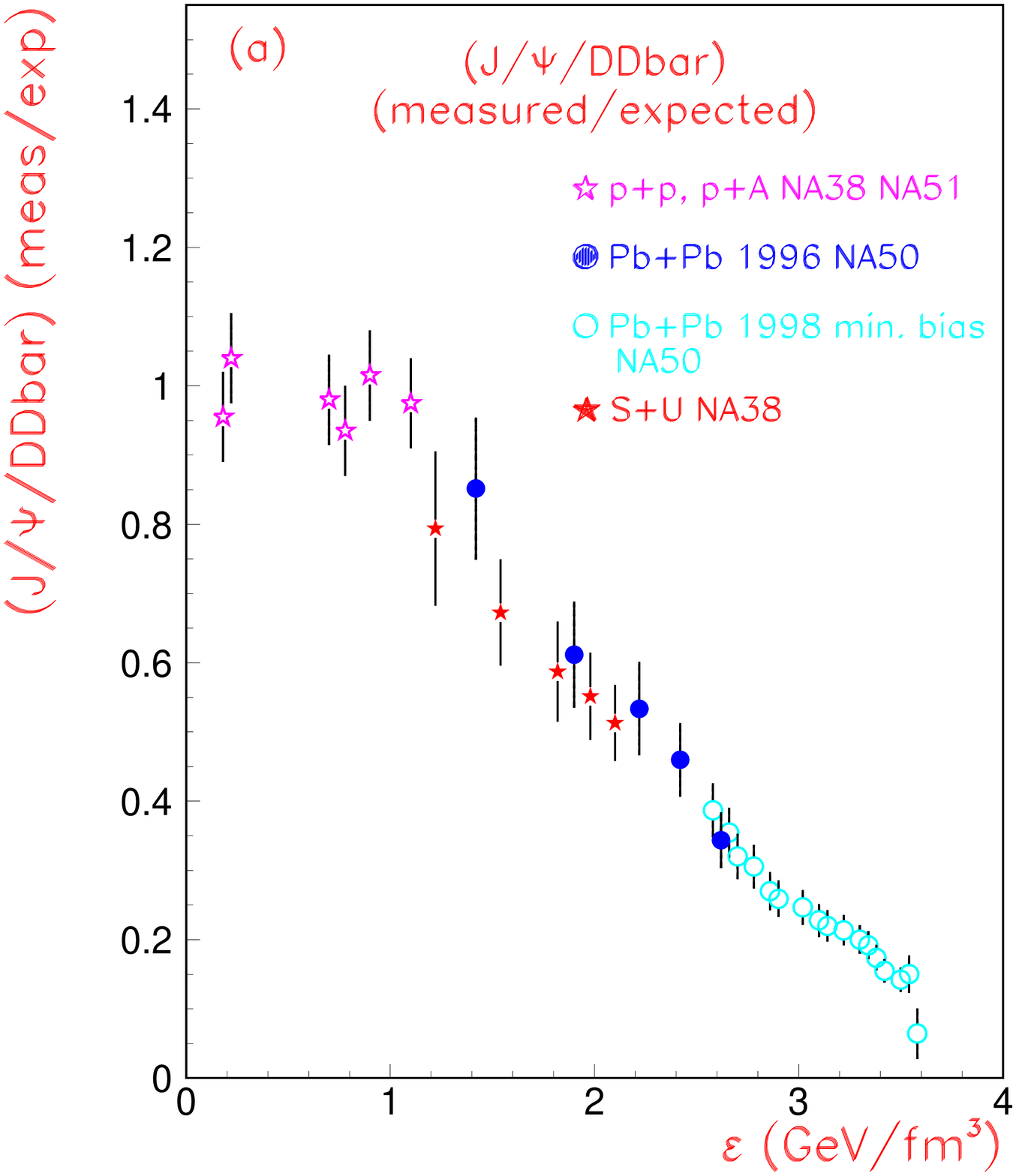,height=3.2in}
\end{center}
\caption{
Left:
The kaon ($\sim$ $K^+$) multiplicity
 over the effective volume of the particle source,
is shown as a function of the initial energy density ($\epsilon$)
\protect\cite{0004138}.
Right: 
 The $J/\Psi/D \overline{D}$ (measured/'expected') ratio is shown
as a function of the initial energy density ($\epsilon$) achieved in the
collisions investigated \protect\cite{0004138}.
\label{fig:k_vs_e}}
\end{figure}

\begin{figure}
\begin{center}
\psfig{figure=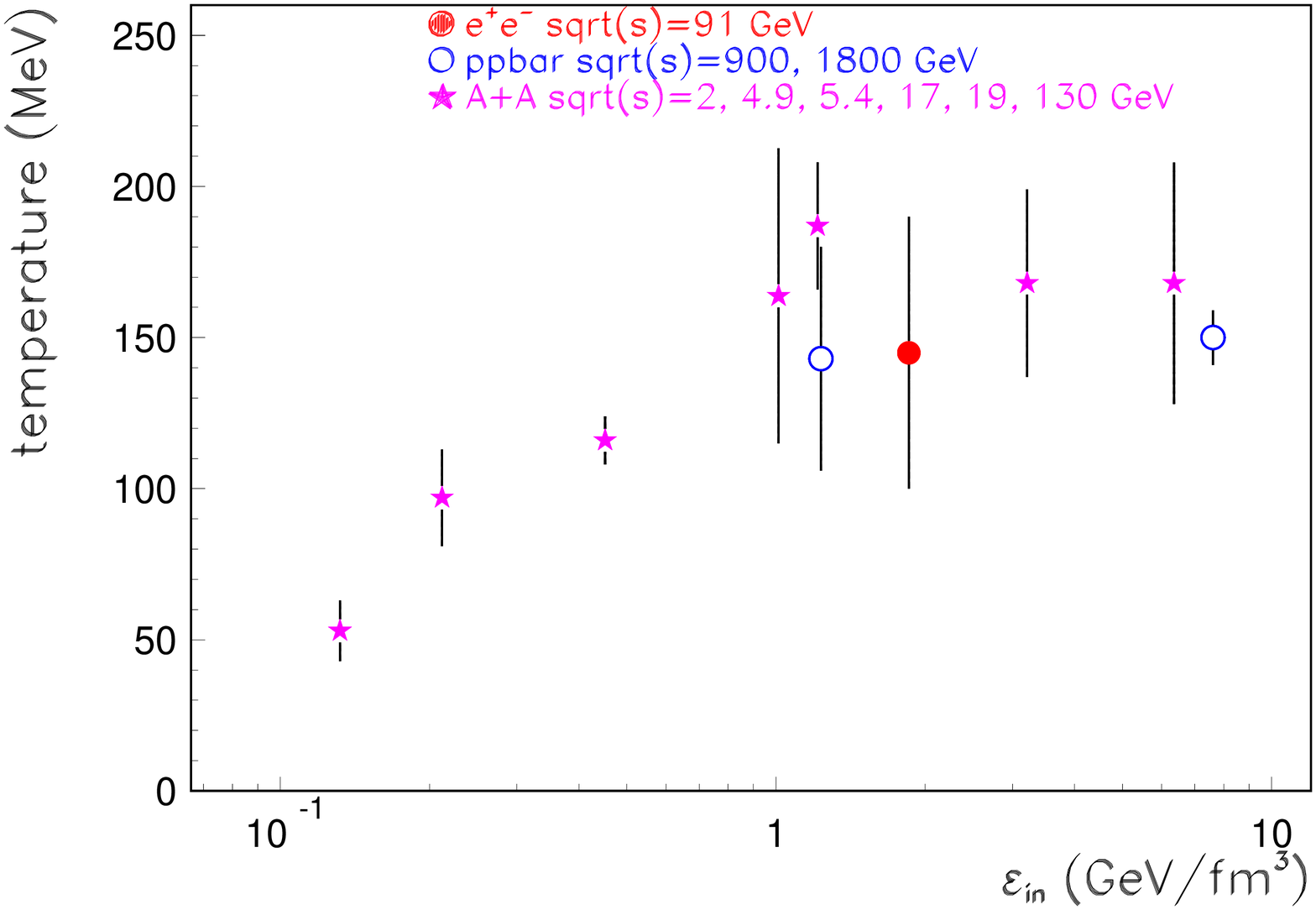,height=2.4in}
\psfig{figure=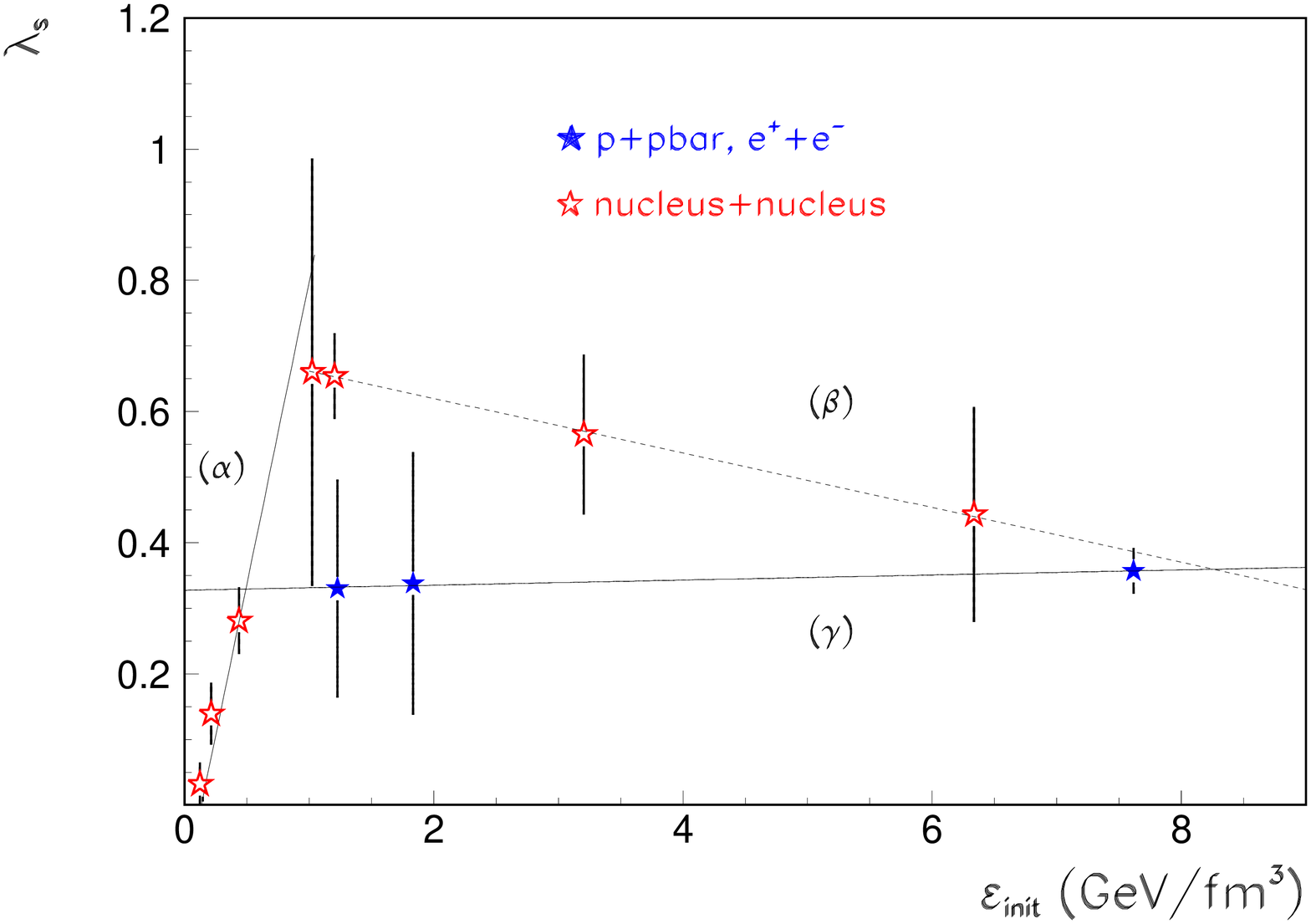,height=2.4in}
\end{center}
\caption{
Left:
The temperature extrapolated to zero $\mu_B$ along
an isentropic path, as a function of
the initial energy density for several A+A, hadron+hadron and 
$e^+e^-$ collisions.
Right: 
The $\lambda_s$ factor as a function of
the initial energy density for several A+A, hadron+hadron and 
$e^+e^-$ collisions.
The lines ($\alpha$), ($\beta$) correspond to 
nonzero $\mu_B$ states, while the line $\gamma$ to
$\mu_B$=0 states.
We demand for the fits confidence level $>$ 10\% \protect\cite{border}.
\label{fig:t_vs_e_mub0}}
\end{figure}

\begin{figure}
\begin{center}
\psfig{figure=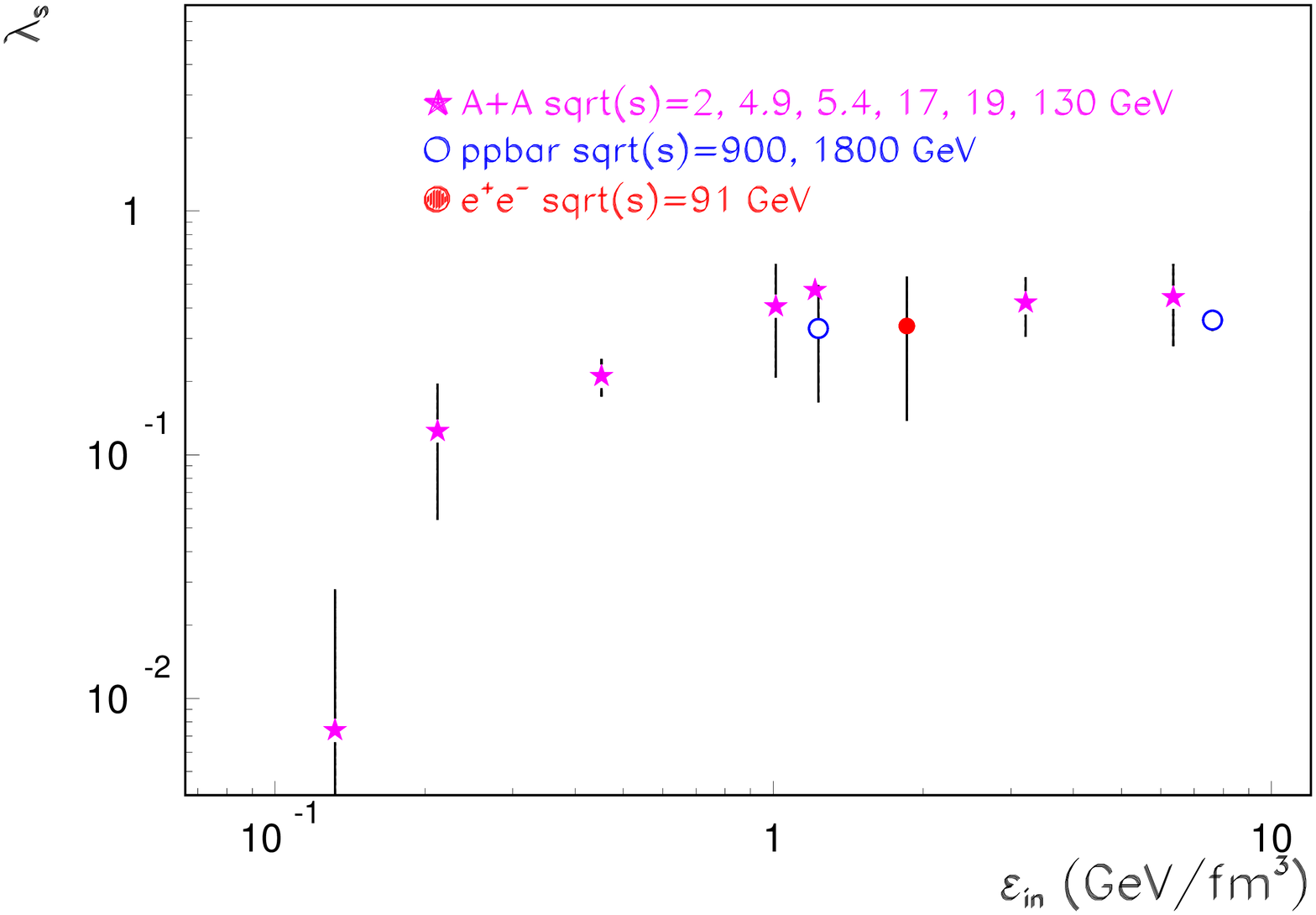,height=2.4in}
\end{center}
\caption{
The $\lambda_s$ factor extrapolated to zero fugacities along
an isentropic path, as a function of
the initial energy density for several nucleus+nucleus, hadron+hadron and
lepton+lepton collisions.
We demand for the fits confidence level $>$ 10\% \protect\cite{border}.
\label{fig:ls_vs_e_mub0}}
\end{figure}

\section{Is there a simultaneous appearance of all QGP signatures ? 
-- Question (a)}

The increase of $m(e^+e^-)$, $m(\mu^+ \mu^-)$ and $\gamma$ 
above expectations in Pb+Pb collisions at 158 
-- and 40 A GeV for $m(e^+e^-)$ \cite{na45_qm2001} --,
is observed above energy density $\epsilon_i$ $\sim$ 1 GeV/fm$^3$.
It is not seen in p+A reactions at 158 GeV/N, which correspond
 to $\epsilon_i$ $<$ 1 GeV/fm$^3$.
\\
The energy density dependence of strangeness production,
 investigated for the first time in \cite{0004138} 
reveals a dramatic change of the kaon number density near 
$\epsilon_i$ = 1.3 GeV/fm$^3$ (figure \ref{fig:k_vs_e} left).
This $\epsilon_i$ coincides with the $\epsilon_i$ at which the
suppression of the $\Psi^{'}/DY$ ratio sets in \cite{na38}, while
the suppression of the $J/\Psi/DY$ ratio starts at a higher $\epsilon_i$
of 2.2 GeV/fm$^3$ \cite{na50}.
\\
It is well possible, that the critical energy density
revealed by strangeness (figure \ref{fig:k_vs_e} left) --and which possibly
shows up also in the $\Psi^{'}/DY$ suppression-- 
is $\sim$ 1 GeV/fm$^3$, and the dissociation of $J/\Psi$ 
occurs at an overcritical $\epsilon_{J/\Psi,dissoc}$ $>$ $\epsilon_{crit}$.
However,
 it is also possible, that all $c \overline{c}$ states
dissociate almost simoultaneously at  $\epsilon_{crit}$.
This follows, assuming that the open charm production is enhanced
in S+U and Pb+Pb collisions at 200 and 158 A GeV, 
following the analysis of NA50 \cite{na50_mm}.
In this case the relevant quantity to search for the $J/\Psi$ suppresion
becomes the $J/\Psi/D \overline{D}$ ratio.
As shown in figure \ref{fig:k_vs_e} right,  this ratio is suppressed 
above $\epsilon_i$ = 1 GeV/fm$^3$, and therefore at the same
$\epsilon_i$ at which strangeness enhancement sets in 
(figure \ref{fig:k_vs_e} left).
\\
We remark that the two behaviour (of $J/\Psi/DY$ and $J/\Psi/D \overline{D}$)
are not compatible.
It appears therefore that a direct measurement \cite{na60} of open charm
is crucial for the interpretation of the $J/\Psi$ suppression pattern.
\\
Another missing piece of information concerning $J/\Psi$, is 
data from $p \overline{p}$ or $pp$ collisions at $\epsilon_i$ between 1 and 3
GeV/fm$^3$, that is, from Tevatron and RHIC.
\\

\section{Is there evidence for the QCD phase transition in global observables ?
\\
Can we extract the critical energy density of QCD from the data ? -- Questions (b) and (c)}

It is a fact, resulting from noumerous thermodynamic analyses
\cite{thermal}, that the various $p+p$ and $A+A$ reactions  studied
 searching for the QCD phase transition,
are described by different baryochemical potentials.
We discuss in the following, what this implies.
\\

\noindent
We follow a Gedankenexperiment, in which we 
identify the water-steam phase transition:
We fill a box with water and look for the
 water-vapour phase transition
 without tools to detect vapour.
 Each time the transition to vapour (=QGP)
occurs
we thus wait until the vapour condensates back to water (=hadron gas),
in order  to measure its temperature.
We make a plot of the water  temperature
as a function of the applied heat, which rises and saturates
at the value of $\sim$ 100$^o$ Celsius.
Adding  salt to water and repeating the experiment would
result in different critical values rising with salinity.
\\
The baryochemical potential is like salt for hadronic systems.
To achieve measuring one single curve one has to use
the same salinity.
Therefore, we arrive at the conclusion, that when extrapolating
all measured $A+A$ and $p+p$ systems to the same $\mu_B$ value
(the simplest one being zero),
 the border of the QCD phase transition can be drawn
and the critical energy density can be extracted
selfconsistently from the data,
independent of a model which  predicts where the boundary
must be \cite{0010228}.
\\

\noindent
We perform this extrapolation in a global analysis of
$e^+e^-$, $p+p$, $p+ \overline{p}$ and
$A+A$ collisions at $\sqrt{s}$ from 2.6 to 1800 GeV per N+N pair
\cite{mapping,border}.
The extrapolation is performed along paths of equal 1) density
2) energy density and 3) entropy density. All three methods give similar
results.
\\
The resulting $T$ at chemical freeze out and at zero $\mu_B$
indeed rises and saturates for all systems which reach
$\epsilon_i$ $\geq$ 1 GeV/fm$^3$, as shown in figure \ref{fig:t_vs_e_mub0} left.
\\
We interpret this behaviour as universally mapping out the 
QCD phase transition in particle and nuclear reactions
  \cite{0010228,mapping,border,jets}.
The critical energy density can therefore be extracted
from the data: $\epsilon_{crit}$ = 1 $\pm$ 0.3 GeV/fm$^3$.
\\
We therefore arrive at a new interpretation of the QCD
phase transition:
\\
1) the discriminating parameter is the initially reached
energy density
\\
2) all reactions, which agree with a thermodynamic description and
  reach $\epsilon_i$ $>$ $\epsilon_{crit}$,
went through the phase transition and back.
\\
3) No 'critical  volume' appears to be needed, which would exclude e.g.
$p \overline{p}$ collisions from going through the phase transition,
once $\epsilon_{crit}$ is reached.
\\
However, as far as the 'critical  volume' is concerned,
the opposite conclusion has been reached in the literature
based on the increase of strangeness
 in $A+A$ as compared to $p+p$ collisions at the same $\sqrt{s}$ 
\cite{lambdas}.
Why this is not so, is the subject of the following section.

\section{A reinterpretation of strangeness as QGP signature -- Questions (d) and (e)}\label{subsec:ssbar}

\noindent
What is the energy density dependence of a global measure
of strangeness, rather than only the kaon number density
(seen in figure \ref{fig:k_vs_e} left) ?
We show in figure \ref{fig:t_vs_e_mub0} right, that 
the 'strangeness suppression factor'
$\lambda_s$ in A+A collisions (all open points), 
increases until $\epsilon_i$ = 1 GeV/fm$^3$,
and then drops. 
The $\lambda_s$ in $p + \overline{p}$ collisions 
(all closed points), defines the $\mu_B$=0 line in this plot.
\\
It appears from this figure  that there is no
sudden drop towards the higher $\epsilon_i$
and saturation of $\lambda_s$ in A+A collisions
right after $\sqrt{s}$=40 GeV, as discussed in \cite{marek},
but a continuous decrease, until the $\mu_B=0$ limit
of $\lambda_s$ is reached at $\epsilon_i$ = 8-9 GeV/fm$^3$,
probably within the reach of the LHC \cite{border}.
\\
This fact becomes visible, when plotting $\lambda_s$ 
as a function of $\epsilon_i$, instead of $\sqrt{s}$.
This is an important change, as e.g. central p+p, S+S and Pb+Pb
collisions at the same $\sqrt{s}$, will all reach
very different $\epsilon_i$.

\noindent
As shown in figure \ref{fig:ls_vs_e_mub0} $\lambda_s$ at $\mu_B$=0,
exhibits a universal behaviour similar to the temperature,
rising until $\epsilon_i$ $\sim$ 1 GeV/fm$^3$ and saturating above,
for all reactions.
It appears from this figure that 
 strangeness is not significantly enhanced in A+A
collisions as compared to $p+\overline{p}$ collisions at the same
$\epsilon_i$. That is, no 'critical volume' is needed.
\footnote{
The somewhat lower $\lambda_s$ values in  $p+\overline{p}$ collisions
as compared to A+A collisions, reflect the lower temperature of the 
$p+\overline{p}$ collisions as seen in figure \ref{fig:t_vs_e_mub0}.
Why is the temperature systematically lower in the
$p+\overline{p}$ collisions ?
One possible explanation is obtained, when visualizing the paths
of the different systems in the $T$, $\mu_b$ plane, while going from
the plasma phase towards their hadronic freeze out:
All systems with nonzero but relatively small 
baryochemical potentials
during hadronization, isentropic expansion and cooling after the transition,
 follow a curved
 path leading them systematically to higher $\mu_B$ and higher $T$,
as compared to a straight path down at the same $\mu_B$.
On the other side all systems with zero $\mu_B$ will hadronize
and cool until their freeze-out, 
following a straight path down, allways at zero $\mu_B$.
It therefore appears plausible, that all systems with nonzero $\mu_B$
will end up at the chemical freeze out with a higher temperature
as compared to the systems with zero $\mu_B$, 
provided they do not stop exactly on the transition curve \cite{workinprogress}.
}

Strangeness is however enhanced in all systems reaching and exceeding
the $\epsilon_{crit}$ of $\sim$ 1 GeV/fm$^3$ as compared to the ones
which are below,  following the temperature.
Within this new interpretation of strangeness as QGP signature,
 recent puzzles discussed in the introduction, can be naturally
explained.
In particular 
\\
a) the increase of the double ratio $K/\pi$(A+A/p+p)
with decreasing $\sqrt{s}$ \cite{ogilvie}, 
\\
b) the maximum of $K^+/\pi^+$
\cite{na49} and of $\lambda_s$ (figure \ref{fig:t_vs_e_mub0} right) at 40 A GeV
Pb+Pb collisions,
\\
c) the difference in the $\sqrt{s}$ behaviour of $K^+/\pi^+$ and
$K^-/\pi^-$ \cite{na49}
\\
d) the difference between the $\sqrt{s}$ behaviour of $K^+/\pi^+$
at midrapidity and in full acceptance \cite{redlich_qm2001}
and 
\\
e) the strangeness enhancement seen partly in p+A data
as compared to p+p \cite{phi},
can be understood as due to the different $\mu_B$ of the 
compared reaction systems, and to the fact that the comparison at the same
$\sqrt{s}$ for different particles, is not at the same $\epsilon_i$.
In
 the particular case of the
$\phi/\pi$ ratio in forward rapidity \cite{phi}, (point (e))
the rapidity dependence of $\mu_B$ is also relevant. 
As point (d) is concerned, the $\mu_B$ at midrapidity is minimal, therefore
the $K^+/\pi^+$ ratio at $y_{cm}$, does not show the drop
seen in the full acceptance $K^+/\pi^+$  versus $\sqrt{s}$.

\section{Conclusions}

\noindent
We present a new interpretation of the QCD phase transition 
and of strangeness as QGP signature.
The starting point of this analysis is the extraction
of thermodynamic parameters describing the final state of 
$e^+ e^-$, $p+p$, $p + \overline{p}$ and $A+A$
reactions between $\sqrt{s}$ = 2.6 and 1800 GeV per N+N pair
\cite{0010228,mapping,border}.
The main new idea is to 
 extrapolate these parameters to zero chemical potentials
 \cite{0010228,mapping,border}.
We then study the temperature and the
 'strangeness suppression factor' $\lambda_s$ ($\lambda_s$=
 $\frac {2 \overline{s} } { \overline{u} + \overline{d} }$)
as a function of the energy density
reached early in each collision (initial energy density $\epsilon_i$).

\noindent
We arrive at the following conclusions:
\\
1) 
After extrapolation to $\mu_B$=0
the temperature rises with $\epsilon_i$, and saturates
above $\epsilon_i$ $\sim$ 1 GeV/fm$^3$.
We interpret this behaviour as mapping out the QCD 
phase transition universally in particle and nuclear
reactions.
\\
In analogy to a water-steam phase transition, the systems
which reach and/or exceed the critical energy density,
have to hadronize back to the maximum allowed hadronic
temperature and a little below,
therefore exhibiting 
a saturating limiting temperature (in analogy to 100$^o$ C).
The extrapolation to $\mu_B$=0, is analogous to
extracting the salt out of the water, in order to
 be able to measure a universal $T_c$.
\\
2) As both $p + \overline{p}$ and A+A reactions exhibit this behaviour 
once they exceed $\epsilon_{crit}$,
 it appears that the energy density is the only discriminating
parameter of this transition, and no 'critical volume' 
is needed in order e.g. to achieve thermalization.
\footnote{
It is conceivable that external probes as $c \overline{c}$ color screenig
and jet quenching, have a different dependence on 'critical volume'
and/or on $\sqrt{s}$
than the global observables which are ingredients of the QGP itself, namely
thermalised
$u, \overline{u}, d, \overline{d}, s, \overline{s}$ quarks and gluons.
This possibility will be probed at  SPS, RHIC and LHC.}
\\
3)
Strangeness is not significantly increased in nucleus nucleus collisions
as compared to elementary particle collisions, if
it is compared 1) at the same (zero) chemical potential
and 2) at the same initial energy density.
However, $\lambda_s$ is found to significantly increase
in all systems which reach or exceed $\epsilon_i$ 
$\sim $ 1 GeV/$fm^3$, as compared to all systems which do not.
Strangeness is found to follow closely the temperature,
rising until $\epsilon_i$ $\sim $ 1 GeV/$fm^3$ and saturating
along the border of the QCD phase transition, namely above
1 GeV/$fm^3$.
\\
4)
This allows us to extract in a model independent way
the critical energy density of the QCD phase transition from the data
in particular $\epsilon_{critical}$ = 1 $\pm$ 0.3 GeV/$fm^3$
as well as the limiting T and $\lambda_s$ values \cite{mapping}.
\\
5) The so called 'strangeness suppression',
 namely the decrease of the $K/\pi$ ratio
(or equivalently of $\lambda_s$)
with $\sqrt{s}$,  from its value in Pb+Pb
collisions at 40 A GeV towards 158 A GeV, is explained
as reflecting the varying chemical potentials
of the heavy ion systems.
\\
6)  Several other
experimental observations have the same origin:
\\
$\bullet$ the increase of the double ratio $K/\pi$ (A+A/p+p)
with decreasing $\sqrt{s}$.
\\
$\bullet$ the flatter behaviour of the $K/\pi$ ratio
as a function of $\sqrt{s}$ when extracted at midrapidity
as opposed to the full acceptance. 
\\
$\bullet$ the difference in the $\sqrt{s}$ dependence
of $K^+/\pi^+$ and $K^-/\pi^-$ ratios.
\\
$\bullet$ the strangeness enhancement seen partly in p+A data
as compared to p+p \cite{phi}.

\noindent
The above discussion
 leads to the expectation that A+A as well as p+p collisions at the LHC
are both well above the critical energy density for the QCD phase
transition and should be investigated both in this spirit.

\section*{Acknowledgments}
I wish to thank the Schweizerischer National Fonds for their support.
I also wish to thank  P. Minkowski and K. Pretzl for interesting discussions,
as well as  A. Capella, Y. Dokshitzer,  L. Montanet, 
 B. Pietrzyk, J. Rafelski, K. Redlich,
and D. Ross for interesting discussions during the conference.

\end{document}